\documentclass[
final
]{dmtcs-episciences}

\usepackage{subfigure}
\usepackage{mathrsfs}

\usepackage{amsmath,amssymb,amsfonts,euscript}
\usepackage{algorithmic, algorithm}
\usepackage{graphicx}
\usepackage[ansinew]{inputenc}
\usepackage{graphicx}
\usepackage{color}
\usepackage{mathrsfs}

\newtheorem{prethm}{{\bf Theorem}}

\newenvironment{thm}{\begin{prethm}{\hspace{-0.5
				em}{\bf}}}{\end{prethm}}

\newtheorem{prepro}{{\bf Theorem}}

\newtheorem{preprop}{{\bf Proposition}}

\newtheorem{precor}{{\bf Corollary}}

\newenvironment{cor}{\begin{precor}{\hspace{-0.5
				em}{\bf}}}{\end{precor}}

\newtheorem{preconj}{{\bf Conjecture}}

\newtheorem{predefi}{{\bf Definition}}

\newtheorem{preremark}{{\bf Remark}}

\newtheorem{preexample}{{\bf Fact}}

\newtheorem{prelem}{{\bf Lemma}}

\newtheorem{prelam}{{\bf Lemma}}

\newtheorem{preprob}{{\bf Problem}}

\newtheorem{preproof}{{\bf Proof}}

\newtheorem{preali}{{\bf Proof of Theorem 1.}}

\newenvironment{ali}[1]{\begin{preali}{\rm
			#1}\hfill{$\Box$}}{\end{preali}}

\newtheorem{prealii}{{\bf Proof of Theorem 2.}}

\newenvironment{alii}[1]{\begin{prealii}{\rm
			#1}\hfill{$\Box$}}{\end{prealii}}

\newtheorem{prealiii}{{\bf Proof of Theorem 3.}}

\newenvironment{aliii}[1]{\begin{prealiii}{\rm
			#1}\hfill{$\Box$}}{\end{prealiii}}

\newtheorem{prealiiii}{{\bf Proof of Theorem 5.}}

\newtheorem{prealiiiii}{{\bf Proof of Theorem 6.}}

\author{Arash Ahadi\affiliationmark{1}
	\and Ali Dehghan \affiliationmark{2}
	\thanks{E-mail Addresses: 
		   arash$_{-}$ahadi@mehr.sharif.edu (Arash Ahadi) $\mathsf{alidehghan@sce.carleton.ca}$ (Ali Dehghan).}}
\title[$(2/2/3)$-SAT problem and its applications in dominating set problems]{$(2/2/3)$-SAT problem and its applications in dominating set problems}

\affiliation{
	Department of
	Mathematical Sciences, Sharif University of Technology, Tehran,
	Iran\\
	Systems and Computer Engineering Department, Carleton University, Ottawa,   Canada
}

\received{2016-5-5}

\revised{2019-3-7}

\accepted{2019-8-5}
\begin{document}
\publicationdetails{21}{2019}{4}{9}{1464}
\maketitle
\begin{abstract}{
The satisfiability problem is known to be $\mathbf{NP}$-complete in general and for many restricted
cases. One way to restrict instances of $k$-SAT is to limit the number of times a variable can be occurred. It was shown that for an instance of 4-SAT with the property that every variable appears in exactly 4 clauses (2 times negated and 2 times not negated), determining whether there is an assignment for variables  such that every clause contains exactly two true variables and two false variables is $\mathbf{NP}$-complete. In this work, we show that deciding the satisfiability of 3-SAT with the property that every variable appears in exactly four clauses (two times negated and two times not negated), and each clause contains at least two distinct variables is $ \mathbf{NP} $-complete. We call this problem $(2/2/3)$-SAT. For an  $r$-regular graph $G = (V,E)$ with $r\geq 3$,  it was asked in [Discrete Appl. Math., 160(15):2142--2146, 2012] to determine whether for a given independent set $T $  there is an independent dominating set $D$ that dominates $T$ such that $  T  \cap D =\varnothing $?
As an application of $(2/2/3)$-SAT problem  we show that  for every $r\geq 3$, this problem is  $ \mathbf{NP} $-complete.
Among other results, we study the relationship between 1-perfect codes and the incidence coloring of graphs and as  another application of our complexity results,
we  prove
that for a given cubic graph $G$ deciding whether $G$ is 4-incidence colorable  is $ \mathbf{NP} $-complete.
}\end{abstract}

\section{Introduction}

The satisfiability problem is known to be $ \mathbf{NP} $-complete in general and for many restricted
cases, for example see \cite{MR3544062,  MR3386014, MR3864719, MR2184613, MR3810276, MR2500722}. Finding the strongest possible restrictions under which the satisfiability problem remains $ \mathbf{NP} $-complete
is important since this can make it easier to prove the $ \mathbf{NP} $-completeness
of new problems by allowing easier reductions.
An instance of $k$-SAT is a set of clauses that are disjunctions of exactly $k$ literals. The problem is to determine
whether there is an assignment of truth values to the variables such that all the clauses are satisfied.
One way to restrict instances of $k$-SAT is to limit the number of times a variable can be occurred.
Consider a 4-SAT formula with the property that each clause contains four variables and each variable appears four times in the formula, twice negated and twice not negated,
determining whether there is a truth assignment for the formula such that in each clause there are exactly two true literals is $\mathbf{NP}$-complete. In this work, we show that a similar version of this problem is $\mathbf{NP}$-complete \cite{Puzzle}.

\subsection{$(2/2/3)$-SAT problem}

For a given formula $\Phi= (X,C)$  a truth assignment is a mapping which
assigns to each variable one of the two values $true$ or $false$.
A truth assignment satisfies a clause $c \in C$ if $c$ contains at
least one literal whose value is $true$.
A truth assignment satisfies a CNF formula (a
Boolean formula in Conjunctive Normal Form) if it satisfies
each of its clauses. Given a CNF formula $F$, the satisfiability
problem asks to determine if there is a truth assignment
satisfying $F$.
We show that deciding the satisfiability of 3-SAT with the property that every variable appears in exactly four clauses (two times negated and two times not negated), is $ \mathbf{NP} $-complete. We call this problem $(2/2/3)$-SAT.
\\ \\
{\em $(2/2/3)$-SAT problem.}\\
\textsc{Instance}: A 3-SAT formula $\Phi=(X,C)$ such that  every variable appears in exactly four clauses (two times negated and two times not negated),  also each clause contains at least two distinct variables. \\
\textsc{Question}: Is there a truth assignment for the variables $X$ of formula $\Phi$ such that each clause in $C$ has at least one  true literal?
\\ \\
Note that if we consider  3-SAT problem  with the property that every variable appears in two clauses (one time positive and one time negative), then the problem is always  satisfiable.
Also, Tovey in \cite{tovey1984simplified} showed that
instances of 3-SAT in which every variable occurs three times are always satisfiable (this is an immediate corollary
of Hall's Theorem).
Also, the following similar restriction is mentioned in the book Computational Complexity by Papadimitriou (page
183): ``Allowing clauses of size two and three with each variable appearing three times and each
literal at most two is  $\mathbf{NP} $-complete (but if all clauses have size three it is in $\mathbf{P} $)." Motivated by above results, in this work  we study the computational complexity of $(2/2/3)$-SAT problem.

\subsection{Independent dominating sets}

Suppose that $G=(V,E)$ is a  graph and let $D,T \subseteq V(G) $.
We say $D$ is a dominating set for $T$, if
for every vertex $v\in T$, we have $v\in D$ or there is a vertex $u\in D$ such that $vu\in E(G)$. 
For a given graph $G$ and independent set $T$, finding an independent dominating set $D$ that dominates $T$ has a lot of applications in the concept of dynamic coloring of graphs, see for example \cite{MR2935408, dehhh, MR3679602}. 
Motivated by those applications, 
for an $r$-regular graph $G = (V,E)$,  it was asked in \cite{dehhh} to determine whether for a given independent set $T $, there is an independent dominating set $D$ that dominates $T$ such that $  T  \cap D =\varnothing $?
As an application of $(2/2/3)$-SAT problem we show that  for every $r\geq 3$, this problem is  $ \mathbf{NP} $-complete.

\begin{thm}\label{thm1}\\
	(i) $(2/2/3)$-SAT problem is $ \mathbf{NP} $-complete.\\
	(ii) Let $r\geq 3$ be a fixed integer. Given $(G,T)$, where $G$ is an $r$-regular graph and $T$ is  a  maximal independent set of $G$, it  is $ \mathbf{NP} $-complete to determine whether
	there is an independent dominating set $D$ that dominates $T $ such that $  T \cap D =\varnothing $.
\end{thm}

The vertex set of every graph without isolated vertices can be
partitioned into two dominating sets \cite{ID3}.
For any  $k \geq 3$, Heggernes and Telle   showed that it is $ \mathbf{NP} $-complete to determine whether a graph can
be partitioned into $k$ independent dominating sets \cite{HT}.
It was shown that it is $ \mathbf{NP} $-hard to determine the chromatic index of a given $k$-regular graph for any $k \geq 3$ \cite{MR689264}. Heggernes and Telle  reduced this problem to their problem. For a given $k$-regular graph $G$ they construct a graph $\mathcal{J}_k$ such that the chromatic index of $G$ is $k$ if and only if the vertices of  $\mathcal{J}_k$ can be partitioned into $k$ independent dominating sets.
It was shown in Appendix that the graph $\mathcal{J}_k$ can be partitioned into $k+1$ independent dominating sets. Determining the computational complexity of deciding whether the vertices of a given connected cubic graph $G$ can be partitioned into a number of independent dominating sets is unsolved and has a lot of applications in proving the  $\mathbf{NP} $-hardness results for other problems.
Here, we focus on this problem and present an application.

We show that deciding whether the vertices of a given  graph $G$ can be partitioned into a number of  independent dominating sets is  $ \mathbf{NP} $-complete, even for restricted class of graphs.

\begin{thm}\label{thm4}
	\\
	$(i)$ For a given connected graph $G$ with at most two numbers in its degree set, determining  whether the vertices of $G$ can be partitioned into a number of independent dominating sets is  $ \mathbf{NP} $-complete.\\
	$(ii)$ Determine  whether the vertices of a given  3-regular graph can be partitioned into a number of independent dominating sets is  $ \mathbf{NP} $-complete.
\end{thm}

\subsection{Incidence coloring}

There is a close relationship between 1-perfect codes and the incidence coloring of graphs. We will use this relationship and prove a new complexity result for  the incidence coloring of  cubic graphs.
An {\it incidence} of a graph $G$ is a pair $(v, e)$ with $v \in V(G)$, $e \in E(G)$, such that $v$ and $e$ are
incident. Two distinct incidences $(v, e)$ and $(w, f )$ are adjacent if one of the following holds:\\
$(i)$ $v = w$, or\\
$(ii)$ $v$ and $w$ are adjacent and $vw \in\{e, f \}$.\\
An {\it incidence coloring} of a graph
$G$ is a mapping from the set of incidences to a color set such that adjacent incidences of
$G$ are assigned distinct colors. The {\it incidence chromatic number} is the minimum number of
colors needed and denoted by $\chi_i(G)$.

The concept of incidence coloring was first introduced by Brualdi and   Massey  in 1993 \cite{Kh2}. They said that
determining the incidence chromatic number of a given  cubic graph is an interesting question. After that the incidence coloring of cubic graphs were investigated by several authors \cite{Kh5, MR3636883, Kh4, Kh6}.
In 2005 Maydanskiy proved that the Incidence Coloring Conjecture\footnote{The incidence coloring conjecture  states that any graph can be incidence-colored with
	$\Delta+2$ colors, where $\Delta$ is the maximum degree of the graph.} holds for any graph with
$\Delta(G)\leq 3$ \cite{Kh4}. Therefore, for  a given cubic graph $G$, $\Delta(G)+1 \leq \chi_i(G) \leq \Delta(G)+2 $.
For a graph $G$ with $\Delta(G)=3$, if the degree of any vertex of $G$ is 1 or 3, then the graph $G$ is called a semi-cubic
graph. In 2008, it was shown that it is $ \mathbf{NP} $-complete to determine if a semi-cubic graph is $4$-incidence colorable \cite{Kh5}. 
Furthermore, recently Janczewski {\it et al.} proved that the incidence 4-coloring problem for semi-cubic bipartite graphs is $ \mathbf{NP} $-complete \cite{janczewski2017incidence}. 
Here, by using the relationship between 
incidence coloring of graphs and independent dominating sets,
we improve the previous complexity results  and show the following theorem.

\begin{thm}\label{TF}
	For a given 3-regular graph $G$ deciding whether $G$ is 4-incidence colorable  is $ \mathbf{NP} $-complete.
\end{thm}

\section{Notation}

We follow \cite{MR1567289, MR1367739} for terminology and
notation are not defined here, and we denote $\{1,2,\ldots,n\}$ by $[n]$.
We denote the vertex set and the edge set of
$G$ by $V(G)$ and $E(G)$, respectively. The maximum degree
and minimum degree of $G$ are denoted by $\Delta(G)$ and $\delta(G)$.
Also, for every $v\in V (G)$ and $X \subseteq V(G)$, $d(v)$, $N(v)$ and $N(X)$  denote the degree of $v$,  the neighbor set of $v$
and the set of vertices of $G$ which has a neighbor in $X$, respectively.
We say that a set of vertices are {\it independent} if there is no edge
between these vertices.
The {\it independence number}, $\alpha(G)$, of a graph $G$ is the size of a largest independent set of
$G$.
A {\it clique} in a  graph $G = (V, E)$  is a subset of its vertices such that every two vertices in the subset are connected by an edge.
A {\it dominating set} of a graph $G$ is a subset $D$ of $V(G)$ such that every vertex
not in $D$ is joined to at least one vertex  of $D$.
For $ k\in \mathbb{N} $, a {\it proper edge $k$-coloring} of $G$ is a function $c:
E(G)\rightarrow [k]$, such that if $e,e'\in E(G)$ share a common endpoint,
then $c(e)$ and $c(e')$ are different.
The smallest integer $k$ such that
$G$ has a proper edge $k$-coloring is called the {\it chromatic index} of $G$ and denoted by $\chi '(G)$. By Vizing's theorem   the chromatic index of a graph $G$ is equal to either $ \Delta(G) $ or $ \Delta(G) +1 $ \cite{MR0180505}.

\section{Proofs}

Here, we show that   $(2/2/3)$-SAT problem is $ \mathbf{NP} $-complete. Next, by using  that complexity result we prove that if 
$r\geq 3$ is a fixed integer, then for a given $(G,T)$, where $G$ is an $r$-regular graph and $T$ is  a  maximal independent set of $G$, it  is $ \mathbf{NP} $-complete to determine whether
there is an independent dominating set $D$ that dominates $T $ such that $  T \cap D =\varnothing $.

\begin{ali}{
		It was shown that $2\text{-in-}(2/2/4)$-SAT is $ \mathbf{NP}$-complete \cite{Puzzle}. 
\\ \\
{\em $2\text{-in-}(2/2/4)$-SAT.}\\
\textsc{Instance}: A 4-SAT formula $\Phi=(X,C)$ such that  each variable appears four
times in the formula, twice negated and twice not negated.\\
\textsc{Question}: Is there a truth assignment for the variables $X$ of formula $\Phi$  such that in each clause there are exactly two true
literals?
\\ \\		
We prove the two parts of the theorem together.
		Assume that $r\geq 3$ is a fixed integer. Let $\Phi$ be a given formula in $2\text{-in-}(2/2/4)$-SAT problem. Assume that $\Phi$ has  the set of variables $X$ and the set of clauses $C$.
		We transform the formula $\Phi$ into a
		formula $\Phi '''$  such that in $\Phi '''$
		each variable appears four
		times in the formula, twice negated and twice not negated. Also,
		$\Phi '''$ has a  satisfying assignment  if and only if $\Phi$ has a  satisfying assignment such that in each clause there are exactly two true literals. Next, we transform the formula $\Phi'''$ into  an  $r$-regular  graph $G' $ with a maximal independent set $T$
		such that the graph $G' $ has an independent dominating set $D$ for $T$ if and only if the formula $\Phi '''$ has a  satisfying assignment.
		Our proof consists of five steps.\\
		{{\bf Step 1.}}\\
		For every clause $c=(\alpha \vee \beta \vee \gamma \vee \zeta)$ in $\Phi$, consider the ten clauses $(\bigstar\alpha \vee \bigstar\beta \vee \bigstar\gamma \vee \bigstar\zeta)$ in $\Phi '$ (for each variable $\alpha$, $\bigstar \alpha$ means one of $\alpha$ or $\neg \alpha$) such that the number of negative literals in each of the clauses in $\Phi '$ is not exactly 2.
		In other words, for instance $(\neg \alpha \vee \neg\beta \vee \gamma \vee \neg\zeta)$ and $ (\alpha \vee \beta \vee \gamma \vee \zeta)$ are in $\Phi '$, but $(\neg \alpha \vee  \beta \vee \gamma \vee \neg\zeta)$ is not in $\Phi '$.
		
		Since in $\Phi $ each variable appears four
		times, twice negated and twice not negated. In $\Phi '$ every literal appears 20 times (each variable appears 40 times). Also, $\Phi $ has a  satisfying assignment such that in every clause there are exactly two true literals if and only if $\Phi '$ has a satisfying assignment (there is at least one true literal in each clause).\\
		{{\bf Step 2.}}\\
		For every clause $c=(\alpha \vee \beta \vee \gamma \vee \zeta)$ in $\Phi'$, consider two new variables $a_c$, $b_c$ and put the following four clauses in $\Phi ''$:
		
		\begin{center}
			$(\alpha \vee a_c \vee b_c )$, $(\beta \vee a_c \vee \neg b_c ), (\gamma \vee \neg a_c \vee b_c ), (\zeta \vee \neg a_c \vee \neg b_c )$.
		\end{center}
		
		It is easy to see that the formula $\Phi '$ has a  satisfying assignment if and only if the formula $\Phi ''$ has a  satisfying assignment.\\
		In the formula $\Phi ''$ some of the variables appear 40
		times. Call them old  variables. For each old variable $x$, consider the new variables $x_0,x_1\ldots,x_{19}, y_0,\ldots,y_{19}$ and for every $i$, $i\in \mathbb{Z}_{20}$, put the following clause in $\Phi ''$:
		
		\begin{center}
			$
			\begin{cases}
			(x_i \vee y_i \vee y_i),\, (\neg x_i \vee \neg y_{i-1} \vee \neg y_{i-1}) ,      &$   if  $   i $ is odd$ \\
			(  x_i \vee \neg y_{i-1} \vee \neg y_{i-1}),\, (\neg x_i \vee y_i \vee y_i)       & $   otherwise. $
			\end{cases}$
		\end{center}
		
		Without loss of generality suppose that the old variable $x$ appears negated in $c_0,c_1, \ldots, c_{19}$ and appears not negated in $c_0',c_1', \ldots, c_{19}'$. For each $i$, $0\leq i \leq 19$ replace $\neg x $ in $c_i$ (respect. $  x $ in $c_i'$) with $\neg x_i$ ( respect. $  x_i$). Call the resulting formula $\Phi '''$. In $\Phi '''$ each variable appears four
		times in the formula, twice negated and twice not negated. It is easy to see that $\Phi ''$ has a  satisfying assignment if and only if $\Phi '''$ has a  satisfying assignment.\\
		{{\bf Step 3.}}\\
		Let $\mathcal{A} =\{A_i : i\in I\}$ be a finite family of (not necessarily distinct) subsets
		of a finite set $\mathcal{U}$. A system of distinct representatives (SDR) for the family $\mathcal{A}$ is a
		set $\{a_i : i \in I\}$ of distinct elements of $\mathcal{U}$ such that $a_i \in A_i$ for all $i \in I$. Hall's Theorem says that $\mathcal{A}$ has a system of distinct representatives if
		and only if $| \cup _{i\in J} A_i|\geq |J|$ for all subsets $J$ of $I$ \cite{MR1367739}.
		
		Let $\Phi'''$ be a given formula with the set of variables $X$ and the set of clauses $C$.
		Let $\mathcal{U}=\{x,\neg x : x\in X\}$ and for every clause $c_i\in C$, $A_i=\{z: z\in c_i\}$.
		In $\Phi '''$ each variable appears four
		times in the formula, twice negated and twice not negated.
		Consider any union of $k$ of the sets $A_i$. Since each $A_i$ contains at least 2 distinct elements and no literal is contained in more than 2 sets, the union contains at least $k$ distinct
		elements. Therefore, by   Hall's Theorem, there exists a system of distinct representatives
		of $\mathcal{U}$. For each clause $c$ denote its representative literal by $z^{c}$. Note that there is a polynomial-time algorithm which finds an SDR, when ever it exists.\\
		{{\bf Step 4.}}\\
		For every variable $x\in X$, put a copy of the complete bipartite graph $K_{r-1,r-1}$ with the vertex set $[\mathcal{X}, \mathcal{Y}]$, where $\mathcal{X}=\{x_i | i\in [r-1]\}$ and $\mathcal{Y}=\{\neg x_i | i\in [r-1]\}$.
		Also, for every clause $c=(x \vee y \vee w)$ put the vertex $c$.
		Join the vertex $c$ to one of the vertices $x_1$ or $x_2$. Also, join the vertex $c$ to one of the vertices $ y_1$ or $y_2$ and join the vertex $c$ to one of the vertices $w_1$ or $w_2$, such that in the resulting graph $\max\{d(x_1),d(x_2),d(y_1),d(y_2),d(w_1),d(w_2)\} \leq r$.
		
		Next, for every clause $c$ join the vertex $c$ to the vertices $z^{c}_3, z^{c}_4, \ldots, z^{c}_{r-1}$.
		Call the resulting graph $G$. In $G$ the degree of each clause vertex $c$ is $r$ and $\delta(G)\geq r-1$.\\
		{{\bf Step 5.}}\\
		Let $K_{r+1}$ be a complete graph with the vertices $v_1,v_2,\ldots,v_{r+1}$. Let $e=v_1v_2$ and $H_r=K_{r+1}\setminus \{e\}$. Consider two copies of $G$. For every vertex $v\in V(G)$ with $d(v)<r$, put a copy of $H_r$ and join the vertex $v_1$ to the vertex $v$ of the first copy of the graph $G$, and join the vertex $v_2$  to the vertex $v$ of the second copy of the graph $G$.
		Call the resulting $r$-regular graph $G'$.
		
		In the following we introduce the members of maximal independent set $T$.
		
		{{\bf  Members of $T$.}}\\
		{{\bf Step 1.}}\\
		For every subgraph   $G$ of $G'$ put the set of vertices $\{c: c\in C\}$ in $T$.\\
		{{\bf Step 2.}}\\
		For every subgraph  $H_r$ of $G'$ put the vertices $v_1$ and $v_2$ in $T$. ($H_r$ was introduced in Step 5, in the construction of $G'$).
		
		Let $D$ be an independent dominating set for $T$ and
		suppose that $c=(x \vee y \vee w)$ is an arbitrary clause. Without loss of generality suppose that $cx_1,cy_1,cw_1\in E(G')$. By the structure of $G'$ at least one of the vertices of the set $\{x_1,y_1,w_1,z^{c}_3, z^{c}_4, \ldots, z^{c}_{r-1} \}$ is in $D$. On the other hand, for every variable $x\in X$, since we put a copy of complete bipartite graph $K_{r-1,r-1}$ with the vertex set $[\mathcal{X}, \mathcal{Y}]$, where $\mathcal{X}=\{x_i | i\in [r-1]\}$ and $\mathcal{Y}=\{\neg x_i | i\in [r-1]\}$ in $G'$.
		Therefore, if $D$ contains a vertex from  $\mathcal{X}$, then it does not have any vertex from $\mathcal{Y}$ and vice versa.
		First, suppose that $G'$ has an independent dominating set $D$ for $T$.
		Let $\Gamma : X \rightarrow \{true,false \} $ be a function such that $\Gamma(x)=true$ if and only if at least one of the vertices $x_1,\ldots, x_{r-1}$ is in $ D$.

		It is easy to see that
		$\Gamma$ is a   satisfying assignment for $\Phi''$. Next, let  $\Gamma : X \rightarrow \{ true,false\} $ be
		a   satisfying assignment for $\Phi''$. For every $x$, put $x_1, x_{2}$  in $D$ if and only if $\Gamma(x)=true$. It is easy to extend this set into an independent dominating set for $T$. This completes the proof.
}\end{ali}

Here, we prove that deciding whether the vertices of a given  graph $G$ can be partitioned into a number of  independent dominating sets is  $ \mathbf{NP} $-complete, even for connected graphs with at most two numbers in their degree set and  3-regular graphs.

\begin{alii}{
		$(i)$ It was shown that  3-colorability of planar 4-regular
		graphs  is NP-complete \cite{MR573644}. For a given 4-regular
		graph $G$ with $n$ vertices we construct a  graph $\mathcal{H}$ with degree set $\{3n,7n\}$
		such that  the vertices of $\mathcal{H}$ can be partitioned into a number of  independent dominating sets is  if and only if $G$ is 3-colorable. Define:
		\\ \\
		\begin{tabular}{r  l}
			$V(\mathcal{H})=$  & $\{v_{j}^{k}|,\, j\in [n],\, k\in [3],\, v\in V(G)\} \cup \{v'|v\in V(G)\},$
			\\
			$E(\mathcal{H})=$ & $\{ v'v_{j}^{k}| j\in [n],\, k\in [3],\, v\in V(G)\}$\\
			$\cup$&$\{v_{j}^{k} v_{j'}^{k'} | j,j'\in [n],\, k, k' \in [3],\, (j, k)\neq (j', k') ,\, v\in V(G)\}$\\
			$\cup$&$\{u_{j}^{k} v_{j'}^{k} | j,j'\in [n],\, k\in [3],\, vu\in E(G)\}.$
		\end{tabular}
		\\ \\
		First, suppose that $G$ is 3-colorable and let $f:V(G)\rightarrow \{0,1,2\}$ be a proper vertex coloring.
		Consider the following partition for the vertices of $\mathcal{H}$:
		
		$P=\{v'| v\in V(G)\}$,

		$P_{\ell}^{h}=\{ v_{j}^{k}| j=\ell,\, f(v)=(h+k \mod 3)  \}$,
		\\ \\
		where $1 \leq \ell  \leq n,\, 0\leq h \leq 2 $.
		It is easy to see that theses sets are disjoint independent dominating sets for $\mathcal{H}$ and a partition for the vertices of $\mathcal{H}$.

		Now, assume that $G$ is not 3-colorable.
		Let $\mathcal{R}=\{v_{j}^{k}|  j \in [n],\, k\in [3],\, v\in V(G)\}$ and $\mathcal{S}=\{v'|v\in V(G)\}$.
		To the contrary suppose that $T_1,T_2, \ldots, T_z$ is a partition of the vertices of $\mathcal{H}$ and each $T_i$ is an independent dominating set for $\mathcal{H}$.
		Consider the following partition for the vertices of $\mathcal{H}$:

		$V(\mathcal{H})=\bigcup_{v\in V(G)}\mathcal{A}_v$.
		
		$\mathcal{A}_v=\{v', v_{j}^{k}| j\in [n],\, k\in [3]\}$.
		\\ \\
		By the structure of $\mathcal{H}$ for every independent dominating set $T_i$ and $v\in V(G)$ we have $ |T_i \cap \mathcal{A}_v |=1$.
		Therefore, for each $T_i$, $|T_i \cap \mathcal{R}|\leq n$, so $z\geq |\mathcal{R}|/n=3n^2/n$.
		Therefore $z  \geq 3n $.
		On the other hand, since $G$ is not 3-colorable, therefore for every independent dominating set $T_i$, we have $|T_i \cap \mathcal{S} |\geq 1$. Consequently $|\mathcal{S}|\geq 3n$. But this is a contradiction. This completes the proof.
		\\
		\\
		$(ii)$  Given a graph $G$,
		a subset $A$ of its vertex set is a 1-perfect code if
		$A$ is an independent set and every vertex not in $A$ is at distance one from exactly one vertex of $A$.
		In other words:
		
		\begin{center}
			$C \subseteq V(G) $ is 1-perfect code $\Leftrightarrow (\forall v\in V(G))(\exists! c\in C)d(v,c)\leq 1 $
		\end{center}
		
		It was shown that for a given 3-regular graph $G$ determining whether the vertices of $G$ can be partitioned into l-perfect codes is $ \mathbf{NP}$-complete \cite{Kh}.
		Note that every l-perfect code in a 3-regular graph on $n$ vertices has size $n/4$.
		So the vertices of a given 3-regular graph $G$ can be partitioned into l-perfect codes if and only if the vertices of $G$ can be assigned 4 different colors in such a way that closed neighborhood of each vertex is
		assigned all 4 colors, i.e.,  $\chi(G^2)=4$ (The square of a graph $G$, denoted by
		$G^2$, is the graph obtained from $G$ by adding a new edge joining each pair of vertices at
		distance 2). Therefore, from \cite{Kh} we have the following corollary:
		
		\begin{cor}
			For a given 3-regular graph $G$ determining whether  $\chi(G^2)=4$ is $ \mathbf{NP}$-complete.
		\end{cor}

		Let $G$ be a 3-regular graph. Let $H = G \cup K_4$. Then the vertices of $G$ can be partitioned into l-perfect codes  if and only if the vertices of $H$ can be partitioned into independent dominating sets. This completes the proof.
}\end{alii}

Next, we show that for a given 3-regular graph $G$ deciding whether $G$ is 4-incidence colorable  is $ \mathbf{NP} $-complete.

\begin{aliii}{
		An {\it strong vertex coloring} of graph $G$ is a proper vertex coloring of $G$ such that for any $u,w\in N_G(v)$, $u$ and $w$ are
		assigned distinct colors. If $c: V (G)\rightarrow S$ is an strong vertex coloring of $G$ and $|S| = k$, then $G$ is called {\it $k$-strong-vertex colorable} and $c$ is a {\it $k$-strong-vertex coloring} of $G$, where $S$ is a color set.
		
		It was shown that for a given graph $G$ whose vertices have degree equal to $k$ or 1 is $(k + 1)$-incidence
		colorable if and only if $G$ is $(k + 1)$-strong-vertex colorable \cite{Kh5}. Since for a given 3-regular graph $G$ determining whether  $\chi(G^2)=4$ is $ \mathbf{NP}$-complete (for more details, see Part $(ii)$ in the proof of Theorem \ref{thm4}), thus for a given 3-regular graph $G$ deciding whether $G$ is 4-incidence colorable is $ \mathbf{NP}$-complete.
}\end{aliii}

\section{Conclusion and Future Works}

\subsection{$(2/2/3)$-SAT problem}

In this work, we proved that  deciding the satisfiability of 3-SAT with the property that every variable appears in exactly 4 clauses (2 time negated and 2 times not negated) and each clause contains at least two distinct variables is $ \mathbf{NP} $-complete. We called this problem $(2/2/3)$-SAT.
Note that if we consider  3-SAT problem  with the property that every variable appears in 2 clauses (1 time positive and 1 time negative), then the problem is always  satisfiable.
Also, Tovey in  \cite{tovey1984simplified} showed that
instances of 3-SAT in which every variable occurs three times are always satisfiable.
It is interesting to determine the complexity of  $(2/2/3)$-SAT problem when each clause has exactly three distinct variables. 
\\ \\
\textbullet $ $   Determine the computational complexity of $(2/2/3)$-SAT problem when each clause has exactly three distinct variables. 

\subsection{Independent dominating sets}
In this work, as an application of $(2/2/3)$-SAT problem, we proved that for each $r\geq 3$, the following problem is $ \mathbf{NP} $-complete: "Given $(G,T)$, where $G$ is an $r$-regular graph and $T$ is  a  maximal independent set of $G$, determine whether
there is an independent dominating set $D$ for $T $ such that $  T \cap D =\varnothing $".
Regarding this result, solving the following question can be interesting.
\\ \\
\textbullet $ $   Determine the computational complexity of deciding whether a given  regular graph has two disjoint independent dominating sets.
\\ \\
We proved that determine  whether the vertices of a given  3-regular graph can be partitioned into a number of independent dominating sets is  $ \mathbf{NP} $-complete. However, one further step does not seem trivial.
\\ \\
\textbullet $  $ Determine the computational complexity of deciding whether  the vertices of a given  connected regular graph can be partitioned into a number of independent dominating sets.
\\ \\
In \cite{dehhh}, it was  proved that
if $G$ is a non-empty graph, and $T$ is an independent set of $G$, then there exists $H$ such that, $H$ is an independent dominating set for $T$ and
$ \frac{\vert T \cap H\vert}{ \vert T\vert}\leq \frac{2\Delta(G) -\delta(G) }{2\Delta(G)}$.
\\ \\
\textbullet $ $   Determine  nontrivial upper bounds for the minimum cardinality of $|I \cap J|$ and also $\frac{|I \cap J|}{I}$ among all  two  independent dominating sets $I$ and $J$ of a graph $G$ for some important family of graphs such as regular graphs.

\subsection{Incidence coloring}

We showed that for a given 3-regular graph $G$ deciding whether $G$ is 4-incidence colorable  is $ \mathbf{NP} $-complete. The complexity of that problem for the family of planar 3-regular graphs can be interesting.

\section{Appendix}
Here, we show that the vertices of $\mathcal{J}_k$ can be partitioned into $k+1$ independent dominating sets.
First, we introduce the construction of $\mathcal{J}_k$. For a given $k$-regular graph $G$, define:
\begin{align*}
V(\mathcal{J}_k)= &  \{v_{e}| v\in V(G), \, e \text{ is incident with }v\}\\
                  &  \cup \{e_{v,u,i}| vu=e\in E(G),\,  i\in [k-1]\}.
\end{align*}
\begin{align*}
 E(\mathcal{J}_k)=  &\{v_e v_{e'}| v\in V(G),\, e,e' \text{ are incident with }v\}\\
                    &\cup \{e_{v,u,i}e_{v,u,i'}| i\neq i'\}\\
                    &\cup \{v_{e} e_{v,u,i}|i\in [k-1],\,  e \text{ is incident with }v \}.
\end{align*}
Now, consider the following useful lemma which will be used in our proof.
\\ \\
{\bf Lemma 1}. Let $G$ be a $k$-regular graph. There is function $f:V\times E\rightarrow [k]$ such that:\\
1. For every edge $e=uv$ in $E(G)$, $f(v,e)\neq f(u,e)$.\\
2. For each vertex $v$ in $V(G)$, for every two edges $e$ and $e'$ incident with $v$, $f(v,e)\neq f(v,e')$.
\\ \\
{\bf Proof of Lemma 1}. Consider the bipartite graph $G^{\frac{1}{2}}$ ($G^{\frac{1}{2}}$ is obtained from $G$ by replacing each edge with a path with
exactly one inner vertex).   Since for every bipartite  graph $H$, $\chi'(H)=\Delta(H)$ (see for example \cite{MR1367739}). Therefore $\chi'(G^{\frac{1}{2}})=k$. Consequently, there is function $f:V\times E\rightarrow [k]$. $\Box$
\\

Partitioning the vertices of a graph into $t$ independent dominating sets is equivalent to a $t$-labeling of the vertices such that each vertex has no neighbors labeled the same as itself and at least one neighbor labeled with each of the other $t-1$ labels.\\
Let $\mathcal{J}_k$ be a graph which is constructed from $G$ and $f$ be a function $f:(v,e)\rightarrow [k]$ such that for every edge $e=uv$ in $E(G)$, $f(v,e)\neq f(u,e)$ and for each vertex $v$ in $V(G)$, for every two edges $e$ and $e'$ incident with $v$, $f(v,e)\neq f(v,e')$.
Consider the following  labeling for the vertices of  $\mathcal{J}_k$:

$\ell(v_{e})=f(v,e)$, $\ell(e_{v,u,1})=k+1$.
\\
For every $e=uv$ define $\ell(e_{v,u,2}), \ldots, \ell(e_{v,u,k-1})$ such that

$\{ \ell(v_{e}), \ell(u_{e}),\ell(e_{v,u,1}), \ldots, \ell(e_{v,u,k-1})\}=[k+1]$.

It is easy to see that $\ell$ is a $(k+1)$-labeling of the vertices such that each vertex has no neighbors labeled the same as itself and at least one neighbor labeled with each of the other $k$ labels. So, the vertices of $\mathcal{J}_k$ can be partitioned  into $k+1$ independent dominating sets. 

\bibliographystyle{plain}
\bibliography{Dynamic}

\end{document}